\documentclass{aa}  

\usepackage{graphicx,xcolor}
\usepackage{txfonts}
\usepackage{natbib}
\begin{document}

\title{A detailed view of filaments and sheets of the warm-hot intergalactic
medium. I. Pancake formation}
\author{J. S. Klar \and J. P. M\"ucket}
\institute{
  Astrophysikalisches Institut Potsdam, An der Sternwarte 16, 14482 Potsdam, Germany \\
  \email{jklar@aip.de, jpmuecket@aip.de}
}
\date{Received 2010, January 12 / Accepted 2010, August 4}
\titlerunning{Filaments and Sheets of the WHIM I: Pancake formation}
\authorrunning{J. S. Klar \& J. P. M\"ucket}

\abstract
% Context
{Numerical simulations predict that a considerable fraction of the missing baryons at
redshift $z \approx 0$ rest in the so-called warm-hot intergalactic medium
(WHIM). The filaments and sheets of the WHIM have high temperatures ($10^5 -
10^7$ K) and a high degree of ionization but only low to intermediate
densities. Therefore, their reliable detection is a challenging task for today's
observational cosmology. The particular physical conditions of the WHIM
structures, e.g. density and temperature profiles, or velocity fields, are expected
to leave their special imprint on spectroscopic observations.}
% Aims
{In order to get further insight into these conditions, we performed hydrodynamical
simulations of the WHIM. Instead of analyzing extensive simulations of cosmological
structure formation, we simulate certain well-defined structures and studied the
impact of different physical processes as well as of the scale dependencies.}
% Method
{We started with a comprehensive study of the one-dimensional collapse (pancake) and
examined the influence of radiative cooling, heating due to an UV background,
and thermal conduction. We investigated the effect of small-scale perturbations given 
by the cosmological power spectrum.}
% Results
{If the initial perturbation length scale $L$ exceeds $\approx 2$ Mpc the
collapse leads to shock-confined structures. As a result of radiative cooling and
of heating due to an UV background a relatively cold and dense core forms in the
one-dimensional case. 
The properties of the core (extension, density, and
temperature) are correlated with $L$. For longer $L$ the core sizes are
more concentrated. Thermal conduction enhances this trend and may even result in
an evaporation of the core. Our estimates predict that a core may start to evaporate for perturbation lengths longer than $L\approx 30$ Mpc. Though the physics in the corresponding three-dimensional case is much more complex, one might expect a similar regulation mechanism with respect to the cold streams along filaments, too. However, this question will be addressed in a forthcoming paper.
The obtained detailed profiles for density and temperature for prototype WHIM
structures allow for the determination of possible spectral signatures by the
WHIM.}
{}


\maketitle

\section{Introduction}

At high redshifts $z \ge 2$ most of the baryons in the Universe rest in the
intergalactic medium (IGM) and can be uniquely described as gas at still low-density 
contrast $\delta \le 10$. It is almost identically distributed as the
underlying dark matter and is highly ionized by the UV background radiation ($T <
10^5$ K). The subsequent evolution changes that picture. At redshift $z \approx 0$,
only a fraction of $\approx 30\%$ of the IGM is still existing under conditions
comparable with those at $z>2$ \citep{Stocke04}. During the evolution toward low
redshifts, the mean-scale streaming motions could have led to shock-confined
filaments containing gas at much higher temperatures.
%+++

Numerical simulations by \citet{CenOstriker99} suggest that approximately 30 \% to
50 \% of the cosmic baryons at $z=0$ are in the form of the intergalactic medium
with a temperature of \mbox{$10^5$ K $<T<$ $10^7$ K}, which is called warm-hot
intergalactic medium (WHIM). Further numerical simulations
\citep[e.g.,][]{Dave01,Dolag06} with different numerical schemes and resolutions
also consistently support this picture. These numerical predictions have initiated
much observational effort in order to reveal the existence of the WHIM.

Owing to the high degree of ionization, the observational signature of the WHIM is very
weak, in particular with respect to neutral hydrogen. Therefore, the detection of
highly ionized metal lines is much more promising. Observationally, the WHIM was
first proposed through its metal absorption features in the spectra of bright
quasars and blazers \citep{Hellsten98,Perna98,Fang00,Cen01,Fang01}. After the first
detection of \ion{O}{vi} absorption lines in the spectra of a bright quasar by \citet{Tripp00} and \citet{Tripp01}, a number of detections were
reported through absorption features of \ion{O}{vi}, \ion{O}{vii}, \ion{O}{viii}, and
\ion{Ne}{ix} ions \citep{Nicastro02,Fang02,Mathur03,Fujimoto04}, but they are
considered to be rather tentative. A detection with sufficiently
high signal-to-noise ratio is reported by \citet{Nicastro05b,Nicastro05a}. They
found absorption signatures of the WHIM at two redshifts in the spectra of the blazar
Mrk421 during its two outburst phases. Future proposed missions such as {\it
International X-Ray observatory (IXO)} are expected to detect numerous WHIM
absorbers. Detection of WHIM absorption in the spectra of afterglows of gamma-ray
bursts (GRBs) were also proposed by \citet{Elvis04} using dedicated missions such
as {\it Pharos} and were considered more recently by \citet{Branchini09} for the
prospects opened by the recently proposed satellite missions \emph{EDGE} and
\emph{XENIA}. \citet{Kawahara06} investigated the feasibility of these detections
in a realistic manner based on cosmological hydrodynamic simulations.

Additionally, several tentative detections of the WHIM through its metal line
emission are claimed by \citet{Kaastra03} and \citet{Finoguenov03} with the \emph{
XMM-Newton} satellite. However, these detections are not significant enough to
exclude the possibility that the observed emission lines are of Galactic origin
because of the limited energy resolution ($\simeq 80$ eV) of the current X-ray
detectors. \citet{Yoshikawa03}, \citet{Yoshikawa04}, and \citet{Fang05} showed
that future X-ray missions equipped with a high-energy resolution spectrograph
such as \emph{DIOS} (Diffuse Intergalactic Oxygen Surveyor) and \emph{MBE}
(Missing Baryon Explorer) can convincingly detect the line emission of the WHIM.

A comprehensive review is given by \citet{ProchaskaTumlinson08}.

It is still an open question how much the WHIM contributes to the anisotropies
of the cosmic microwave background radiation via the Sunyaev-Zeldovich
effect (SZ-effect). Although the density contrast of the WHIM is moderate
($\delta < 100$), its temperature is high ($10^5$ K $< T <$ $10^7$ K), and it is supposed
to make a significant contribution to the cosmic baryon budget of $\approx 50
\%$. Estimates provided by \citet{Atrio06}, \citet{Atrio08}, and
\citet{GenovaSantos09} indicate on a non-negligible contribution, which under
certain conditions might be even comparable with the overall SZ contribution of
clusters of galaxies. Thus the SZ effect could serve as an additional detection
channel for the WHIM. However, the strength of a thermal SZ effect is still a matter of 
debate because the results obtained by numerical simulations are much less
pronounced.

Therefore, it is of principal importance to investigate the detailed
thermodynamic state and the internal kinematics of the structures which may hide a large fraction of cosmic baryons.
The detailed physics of the WHIM is highly demanding for computational
astrophysics, however. The treatment of low-density regions in great detail is difficult.
Compared with the numerical handling of high-density matter distributions where
adaptive techniques can be applied, for low-density regions, necessary higher overall
particle and/or grid number is unavoidable for an appropriate description. In
addition, higher resolution calls for a more detailed consideration of local
physics, e.g., star formation, feedback, contamination by heavy elements, etc.
Altogether the computational effort is immensely more complicated if considered
within a cosmological context \citep{Schaye09,Bertone09}. Still, an adequate
treatment of low-density regions is despite of the currently available highly
developed computational techniques at the limit or still beyond the near future
potential.

In this paper, we consider the formation of WHIM structures at $z=0$.
Contrary to the extensive and complex treatment within the context of
cosmological structure formation, we investigate the evolution of a single one-dimensional
prototype sheet including most of the relevant physical processes. We will determine the detailed temperature and density profiles and determine which
processes the latter might depend on. In a forthcoming paper we will
extend this study toward three-dimensional structures. We will use a similar approach as presented here to represent a halo - filament scenario. This will enable us to obtain further information, in particular about the spatial velocity field.

Although we avoided performing very time-consuming full cosmological
simulations we obtain sufficient reliable information about the pre- and
post-shock evolution of the WHIM structures depending on the assumed initial conditions
(amplitude and scale-size of initial perturbations). The temperature and density
profiles are mainly defined by the latter parameters and the physical processes
that are incorporated. The knowledge of the temperature and density distribution
within WHIM sheets is important for the estimate of the probability to
observe particular features (tracers) in the related spectra. This concerns the
principal observability of certain lines or combinations of lines as well as the
probability for observations (covering factor) in general. 

The paper is organized as follows: In the next section we give the
rationale for the considerations given in the paper and address our basic assumptions. In
Sect. \ref{sTheory} we describe the system of our hydrodynamical equations and basic
relations. In Appendix \ref{aUV}, we provide the system of chemical rate
equations and the approximated model for the UV background evolution. We also shortly
describe the numerical code \texttt{evora}, which we specifically developed for this study. The details are given in Appendix \ref{aCode}. In Sect. \ref{s1D} we investigate the one-dimensional
collapse (pancake formation) for different incorporated physics in detail. The dependence of the results on
the initial perturbation scale is shown in Sect. \ref{sScaling}. We
discuss our results and their possible implications in the final section.

\section{Pancake formation as a model for the WHIM}\label{sPancake}

Most of the IGM gas distribution is the result of one-and two-dimensional
collapse processes. The basic theory for the formation and evolution of structure
is already well understood since the sixties of the last century
\citep{Doroshkevich64}. According to these theories, the most probable formation process
starts first with the one-dimensional collapse. Only if the underlying dark
matter-distribution enters the first caustic, multi-streaming of matter leads to the formation
two-dimensional filaments and eventually knots, which are characterized by matter collapsing
in three dimensions. This evolution is closely followed by the gas
distribution. A basic description of the gas physics including shock
appearance was given in the pioneering work of \citet{Sunyaev72} in the
context of galaxy formation. The directions (orientations) for the
one-dimensional collapsed sheets are determined by the highest eigenvalue of the
deformation tensor, which can be attributed to the initial linear density
perturbations. According to \citet{Doroshkevich78} the probability that two or
even three of the initial eigenvalues are identical or nearly equal is extremely
low. Therefore, the one-dimensional collapse, and at a certain evolutionary stage
the two-dimensional collapse, is the dominating structural evolution process.

The WHIM structures under consideration (sheet or filamentary structure) are
supposed to reach the non-linear stage of evolution not later than at $z=0$. If
the perturbation scale is large enough, the initially perturbed spatial region
remains Jeans-unstable throughout the whole cosmological evolution, i.e.,
when the mean IGM temperature was raised to about $T \approx 10^4$ K during the
cosmic reionization. In order to form shocks, the infall velocity of the collapsing 
gas must reach the speed of sound or even go beyond. These conditions lead us to 
perturbations on scales initially larger than 1-2 Mpc comoving. The structures
arising from those perturbations are expected to form the large-scale network of
the WHIM.

We start our investigation with the consideration of the one-dimensional
collapse of one perturbation with a given length scale. This scenario is
commonly referred to as \emph{cosmic pancake formation}. The particular importance of
studying the one-dimensional planar collapse was stressed by
\citet{Struck-Marcell88}. The advantages to restrict ourselves to the
simplest geometrical structures are obvious:
\begin{itemize}
\item The one-dimensional pancake formation describes the most common collapse
formation process in the universe at large scales ($>$ 1 Mpc).
\item It describes the preliminary phase (predecessor) of collapse processes
of higher dimensions.
\item It allows a spatial resolution far beyond recent capabilities
for three-dimensional simulations.
\item For special cases analytical solutions are available. Those may serve as
test cases \citep[e.g.,][]{Bryan95,Teyssier02}.
\item The high symmetry of the considered configurations does not influence the
physical state and the principal distributions of temperature and density, but
it allows for an additional check for possible numerical instabilities and
deviations of non-physical origin.
\item It allows us to investigate and to control the influence of various,
subsequently introduced, energetic processes onto the thermodynamical
evolution.
\end{itemize}

We include all relevant processes of radiation cooling as well as the
heating by photoionization due to the evolving cosmic UV background. To that
purpose we self-consistently compute the abundances of the different
ionization levels assuming a primordial medium, i.e. consisting of hydrogen (H)
and helium (He) only. Besides using the common assumption of ionization
equilibrium (IE), we perform simulations where the chemical composition is
directly computed by integration of the corresponding differential equations, and
check for differences between the two cases. While in many cases IE is a valid
assumption, deviations from equilibrium may occur, in particular at low
densities. Furthermore we consider the role of thermal conduction. Heat
conduction was also included by \citet{Bond84} considering the
problem of cooling pancakes with analytical methods. At certain conditions very high
temperature gradients are expected to occur. This may happen particularly in
the vicinity of shock fronts. Then, thermal conduction can lead to a considerable
change of the temperature profiles. In the cases to be considered here, the gas
is almost fully ionized and the expression for the heat conduction coefficient
from \citet{Sarazin88} can be used.

The distribution of the gas in the structures of the WHIM (sheets and filaments)
is supposed to be very close to that of the dark matter. This is true even at
late evolution stages. In order to simplify the calculations, we consider the
baryonic content of the Universe only and therefore decrease the number of equations to be solved. Because this would neglect the gravitational mass of the dark matter, we assume that the dark matter obeys the same spatial distribution as the baryons. Concordantly, we rescale the baryonic density to the total cosmic matter density when computing the gravitational potential. 
For test cases, we checked our results for deviations from the solutions including the full
dark matter dynamics. For that purpose, we used the \texttt{RAMSES} code
\citep{Teyssier02} and appropriate initial conditions. The deviation is negligible for the structures
considered here, 

Though considering preferentially one-mode perturbations, we also investigate up to
which degree small-scale perturbations may affect the results. For that purpose, we add
Gaussian random perturbations according to the cosmological initial power-spectrum.
The spatial scale sizes of these fluctuations are lower compared to that of the considered
large-scale single mode, but much higher than the (comoving) initial Jeans length immediately after 
reionization. We will compare the various resulting density and temperature profiles.

\section{Theoretical framework}\label{sTheory}

\subsection{Dynamical equations}\label{sDyn}

We use the standard approach for describing the baryonic component of the
universe. The assumed ideal polytropic fluid is described by the Euler equations,
which may be considered as conservation laws for the \emph{conserved quantities}:
the density $\rho$, the momentum densities $\rho  \mathbf{u}$ ($\mathbf{u}$
denotes the vector of velocities) and the energy density $E$. The latter is the
sum of the kinetic energy density $E_{kin} = 1/2 \, \rho |\mathbf{u}|^2$ and the
internal energy density $E_{th}$. The internal energy density is related to the
pressure $p$ by the polytropic equation of state $p = \left( \gamma - 1 \right)
E_{th}$, where $\gamma$ denotes the adiabatic index of the gas. Throughout this
paper we use the adiabatic coefficient for a mono-atomic gas, i.e., $\gamma
= 5/3$.

In order to include the cosmological expansion into our simulations we use
\emph{supercomoving coordinates} \citep{MartelShapiro98}. This is a
transformation of the physical coordinate $\mathbf{r}$ and the time $t$ into the
supercomoving coordinates $\mathbf{x} = (x,y,z) = \mathbf{r} / a$ and the
conformal time $\textrm{d}\tau = \textrm{d}t / a^2$, where $a$ is the cosmological expansion factor
computed from the Friedman equation. We use a $\Lambda$CDM cosmology with
the parameters derived from the five-year WMAP observations \citep{Komatsu09}
$\Omega_\Lambda = 0.73$, $\Omega_m = 0.27$, and $H_0 = 71$ Mpc km$^{-1}$
s$^{-1}$. The transformed Euler-equations that are used throughout this paper
are
\begin{equation}\label{eRho}
% density
\frac{\partial \rho}{\partial \tau} 
+ \nabla \cdot \left( \rho \mathbf{u} \right) = 0
\end{equation}
\begin{equation}\label{eM} % momenta
\frac{\partial \left( \rho \mathbf{u} \right)}{\partial \tau}
+ \nabla \cdot \left( \rho \mathbf{u}\otimes\mathbf{u} \right) + \nabla p = -
\rho \nabla \phi
\end{equation}
\begin{equation}\label{eE} % energy
\frac{\partial E}{\partial \tau} 
+ \nabla \cdot \left(\mathbf{u} \left( E + p \right) \right) = - \rho \mathbf{u}
\cdot \nabla \phi + \left( \Gamma - \Lambda \right) - \nabla \cdot \mathbf{j}
\end{equation}
\begin{equation}\label{eS} % entropy
\frac{\partial S}{\partial \tau} 
+ \nabla \cdot \left( S \mathbf{u} \right) = \frac{\gamma - 1}{\rho^{\gamma-1}}
\left(\left( \Gamma - \Lambda \right) - \nabla \cdot \mathbf{j}\right) \;.
\end{equation}
These equations already include the change in energy owing to the heating function
$\Gamma$, the cooling function $\Lambda$, and the heat flux $\mathbf{j}$ caused by
thermal conduction. The gravitational potential $\phi$ is computed using the
supercomoving version of Poisson's equation
\begin{equation}\label{ePoisson} % poisson
\Delta \phi = 4 \pi G a \; \frac{\rho_\mathrm{tot} - \bar{\rho}}{\bar{\rho}} \;,
\end{equation}
where $\bar{\rho}$ denotes the uniform background density of the Universe and $\rho_\mathrm{tot}$ the total matter density (baryons + dark matter). As already mentioned, we assume similar spatial distributions for the dark matter and the baryons, and therefore compute the total matter density by $\rho_\mathrm{tot} = \rho / f_B$ assuming a baryon fraction of $f_B = \Omega_b / \Omega_m = 0.16$.

Equation (\ref{eS}) describes the evolution of the modified
entropy density $S = p / \rho^{\gamma- 1}$, which is necessary 
for the dual-energy formalism
described in the Appendix \ref{aHighMach}. It can be derived from Eq.
(\ref{eRho} - \ref{eE}). 
The only difference of Eq. (\ref{eRho} - \ref{ePoisson})
with respect to the non-comoving equations is the 
factor $a$ in Eq. (\ref{ePoisson}) (an additional drag term
would occur in Eq. (\ref{eE}) and Eq. (\ref{eS}) for $\gamma \neq 5/3$).
An extensive derivation of the supercomoving coordinates is given in the
Appendix of \citet{Doumler09}.

In order to follow the chemical network an equation
of continuity for the number densities $n_i$ is needed:
\begin{equation}\label{eN} % number densities
\frac{\partial n_i}{\partial \tau}
+ \nabla \cdot \left( n_i \mathbf{u} \right)
= \Xi_i \;.
\end{equation} 
where $\Xi_i$ denotes the source term due to chemical processes. The index $i$
indicates the five different species \ion{H}{i}, \ion{H}{ii},
\ion{He}{i}, \ion{He}{ii}, and \ion{He}{iii}. The 
electron number density can be computed using charge conservation:
\begin{equation}
n_e = n_\ion{H}{ii} + n_\ion{He}{ii} + 2 \, n_\ion{He}{iii} \;.
\end{equation}
Initially, the number densities can be computed from the density by $n_i = \chi_i
\, \rho / m_i$, where $\chi_i$ denotes the primordial mass fraction of Hydrogen
$\chi_\ion{H}{} = 0.76$ or Helium $\chi_\ion{He}{} = 0.24$, respectively, and
$m_i$ is the corresponding atomic mass.
The temperature necessary for computing the different rates of the chemical
network is given by
\begin{equation}\label{eT}
T = \frac{p}{k_\mathrm{B} \sum_i n_i} \;.
\end{equation}
It is often assumed in cosmological simulations, that the time scales of the
chemical processes are much shorter than the dynamical times. The system is
therefore assumed to be in ionization equilibrium (IE). Then, the
left-hand side of Eq. (\ref{eN}) vanishes and the number densities can be
computed locally using $\Xi_i = 0$. Without that assumption, the chemical source
terms have to be implemented self-consistently using the full set of Eq.
(\ref{eN}). This includes the hydrodynamical advection given by the second term.
In both cases the chemical source term, as well as cooling and heating, are
computed from photoionization, collisional ionization and recombination,
dielectric recombination of \ion{He}{ii}, collisional excitation of \ion{H}{i}
and \ion{He}{ii}, and bremsstrahlung (three-body processes are neglected). The
corresponding rates are taken from \citet{Black81} and \citet{Katz96}. The
details of the chemical network and the model for the UV background are presented
in Appendix \ref{aUV}.

Thermal conduction is implemented as presented in \citet{Jubelgas04}. The heat
flux is computed by $\mathbf{j} = - \kappa(T) \, \nabla T$, where $\kappa$ is the
heat conduction coefficient. We use the coefficient given in \citet{Sarazin88}
derived from the classical thermal conductivity due to electrons from
\citet{Spitzer62}:
\begin{equation}
\kappa = 4.6 \times 10^{13} \, \left(\frac{T}{10^8 \textrm{ K}}\right)^{2.5} 
\left(\frac{\ln\Lambda}{40}\right)^{-1} 
\textrm{erg s}^{-1}\textrm{ cm}^{-1}\textrm{ K}^{-1} \;,
\end{equation}
where $\ln\Lambda$ is the Coulomb logarithm. We use $\ln\Lambda= 37.8$. 
At low densities the mean free path of the electrons $\lambda_e$ 
given by
\begin{equation}
\lambda_e = 0.023 \, \left(\frac{T}{10^8 \textrm{ K}}\right)^{2} 
\left(\frac{n_e}{10^{-3} \textrm{ cm}^{-3}}\right)^{-1}\textrm{Mpc}
\end{equation}
can approach the scale length of the temperature gradient 
$\lambda_T = T /|\nabla T|$. Then the heat flux becomes
saturated. To take this effect into account, we use an effective
coefficient \citep{Sarazin88}
\begin{equation}
\kappa_{\mathrm{eff}} = \frac{\kappa}{1 + 4.2\, \lambda_e \,/ \lambda_T} \;,
\end{equation}
This approach neglects the influence of magnetic fields. They 
mainly affect dense objects like clusters of galaxies.
We focus on regions of low to intermediate density. Therefore this
approximation is sufficient for our study.
 
\subsection{Numerical implementation}\label{sCode}

In order to compute the time evolution of the fluid governed by the equations
introduced in Sect. \ref{sDyn}, we developed the new simulation code
\texttt{evora}. The code uses a regular grid of equally sized cells and evolves
cell-averaged quantities over discrete time steps. Furthermore, the set of
equations is split into four subproblems: hydrodynamic advection, gravitational
acceleration, integration of the chemical network, and thermal conduction. These
problems are solved  successively at every time step, and every solver uses the
quantities updated by its predecessor as its input. Here we give a brief overview
of the four solvers:
\begin{enumerate}
  \item The hydrodynamic problem is solved using the well-known MUSCL scheme in
  combination with the MINMOD slope limiter and the approximate HLLC Riemann
  solver \citep[see][]{Toro99}.
  \item The gravitational potential is computed from Poisson's equation using
  Fourier transforms and Green's function \citep[see][]{Hockney88}.
  \item Under the assumption of IE, the number densities are computed by
  iteration from $\Xi_i = 0$. In the non-IE case they are integrated
  self-consistently using the \emph{modified Patankar scheme}
  \citep{Burchard03}. In both cases the cooling and heating functions are
  applied to the pressure using the regular Patankar scheme
  \citep[see][]{Patankar80}.
  \item Thermal conduction is included into our simulations by using 
  a central difference scheme.
\end{enumerate}
A more detailed description of our code can be found in Appendix \ref{aCode}.

\section{One-dimensional collapse}\label{s1D}

\subsection{Gravitohydrodynamics}

\begin{figure}
\centering
\resizebox{\hsize}{!}{\includegraphics{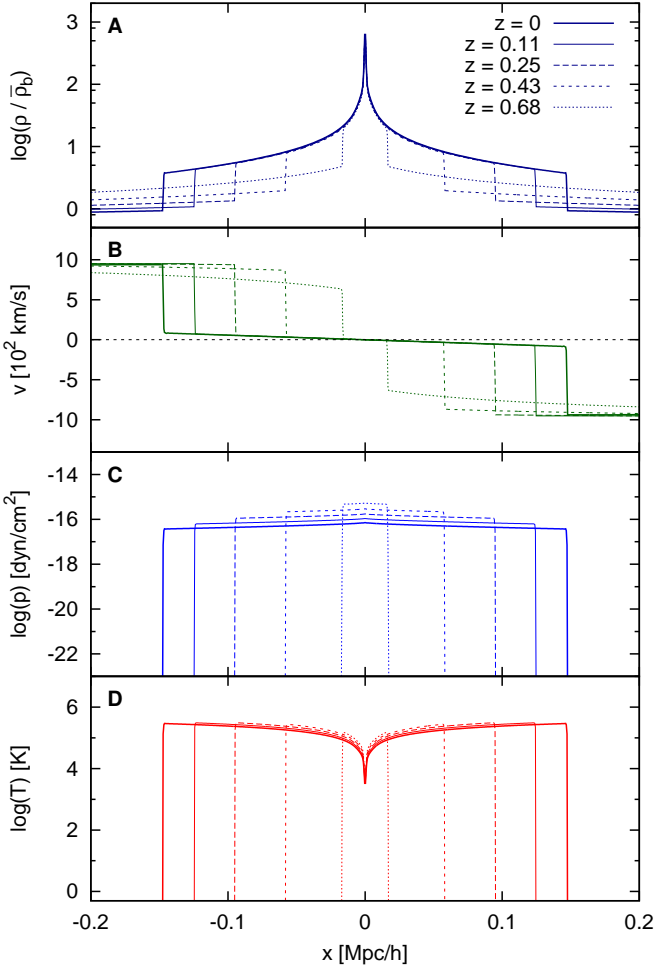}}
\caption{Pancake formation without cooling, heating and thermal
conduction: Profiles for different redshifts. 
\emph{Panel A:} Density; 
\emph{Panel B:} Velocity;
\emph{Panel C:} Pressure; 
\emph{Panel D:} Temperature.
This simulation uses an
initial amplitude of $A_i = 0.02$ at $z=99$, a perturbation scale of $L =
8$ Mpc, and 16000 grid points.}
\label{fPro}
\end{figure}

In a first step we consider the formation of a pancake \emph{without} the
inclusion of any cooling and heating or of thermal conduction. This
serves as a basic configuration to be compared with the results after
successively taking into account relevant physical processes.
The configuration is initialized at $z=99$ by imposing a single sinusoidal
perturbation with an initial amplitude $A$ and a wavenumber $k = 2 \pi / L$
(where $L$ denotes the length scale of the perturbation) with respect to the background
density of the Universe. This density distribution is then scaled to the baryonic density using $f_B$:
\begin{equation}\label{ePancake}
  \rho = \frac{1}{f_B} \, \left( A \, \cos\left( k \, x \right) + \bar{\rho} \right) \;.
\end{equation}
According to the linear perturbation theory \citep[see][]{Peebles93} we obtain the
corresponding velocity field
\begin{equation}
  u = - \frac{f \, \dot{a}}{a} \frac{A \sin( k \, x )}{k} \;,
\end{equation}
where $f= (\mathrm{d} \log D)/(\textrm{d} \log a)$, with $D$ denoting the growth
factor. The initial temperature is set to 100 K. We always choose the size of our
computational domain to be equal to the perturbation scale $L$. For any of our
simulations, periodic boundary conditions are imposed.

Before $z \sim 1$ the medium undergoes adiabatic contraction, resulting in a sharp
density peak in the center of the box. When the local speed of sound matches the
infall velocity, two shocks form and confine a region of high temperature. In
Fig. \ref{fPro} we show density, velocity, pressure, and
temperature profiles at different redshifts after the moment of shock formation.
While moving outward, infalling cold gas passes these shocks and its kinetic
energy is transformed into heat. The associated strong decline in velocity is
visible in the corresponding profile. As a result, the temperature inside
the shocked region is several orders of magnitude higher then outside. This
process is commonly called \emph{shock heating}. While passing the shock,
the gas looses most of its velocity and does not move further inward, but is accumulated at 
the outer wings of the profile, leaving the inner part
unchanged. With time, a continuously decreasing fraction of matter remains
outside of the shocked region. The accretion onto the pancake declines over time. This
results in a slower shock speed and in a declining density profile. The final
density profile bound by the shocks covers about 2.5 orders of magnitude, and is
proportional to $r^{-2/3}$ \citep{Shandarin89}. The pressure profile
remains almost constant. This is an expected behavior since pressure gradients
would be quickly erased by hydrodynamic advection. The small deviation from
uniformity as well as the weak redshift dependence are the imprint of the
gravitational potential and the cosmological expansion. This almost isobaric
behavior can be used to explain the shape of the temperature profiles. Given a
constant pressure, Eq. (\ref{eT}) implies an inverted behavior between
temperature and density. The temperature is given in physical units 
implying a cosmic evolution $\propto a^{-2}$.

Without heating and cooling, the physical dimensions can be eliminated
from the hydrodynamic equations, and therefore the qualitative outcome of these
simulations does not depend on the given length scale $L$. If we impose a
reference system given by $L$, the background density of the Universe
$\bar{\rho}$, and the Hubble time $1/H_0$, we are able to obtain scaling
relations for all quantities. For the temperature scale this yields
\begin{equation}
T_{\mathrm{scale}} = H_0^2 \, L^2 \propto L^2 \;.
\end{equation}
Thus the temperature in the shocked region scales as the square of the length
scale of the initial perturbation. 

Besides the length scale of the perturbation, the initial amplitude is set as a
parameter. Its value determines the time of caustic formation, as shown in
\citet{Bryan95}. The chosen value of $A = 0.02$ corresponds to a
shock formation at redshift $z \sim 1$, which could be a reasonable value for the
WHIM. Owing to the cosmological expansion, the temperature declines very fast from
its initial value. Therefore, the initial temperature has a negligible influence
on the dynamical evolution and on the resulting profiles.

\subsection{Cooling and heating}\label{sCool}

\begin{figure*}
\centering
\includegraphics[width=17cm]{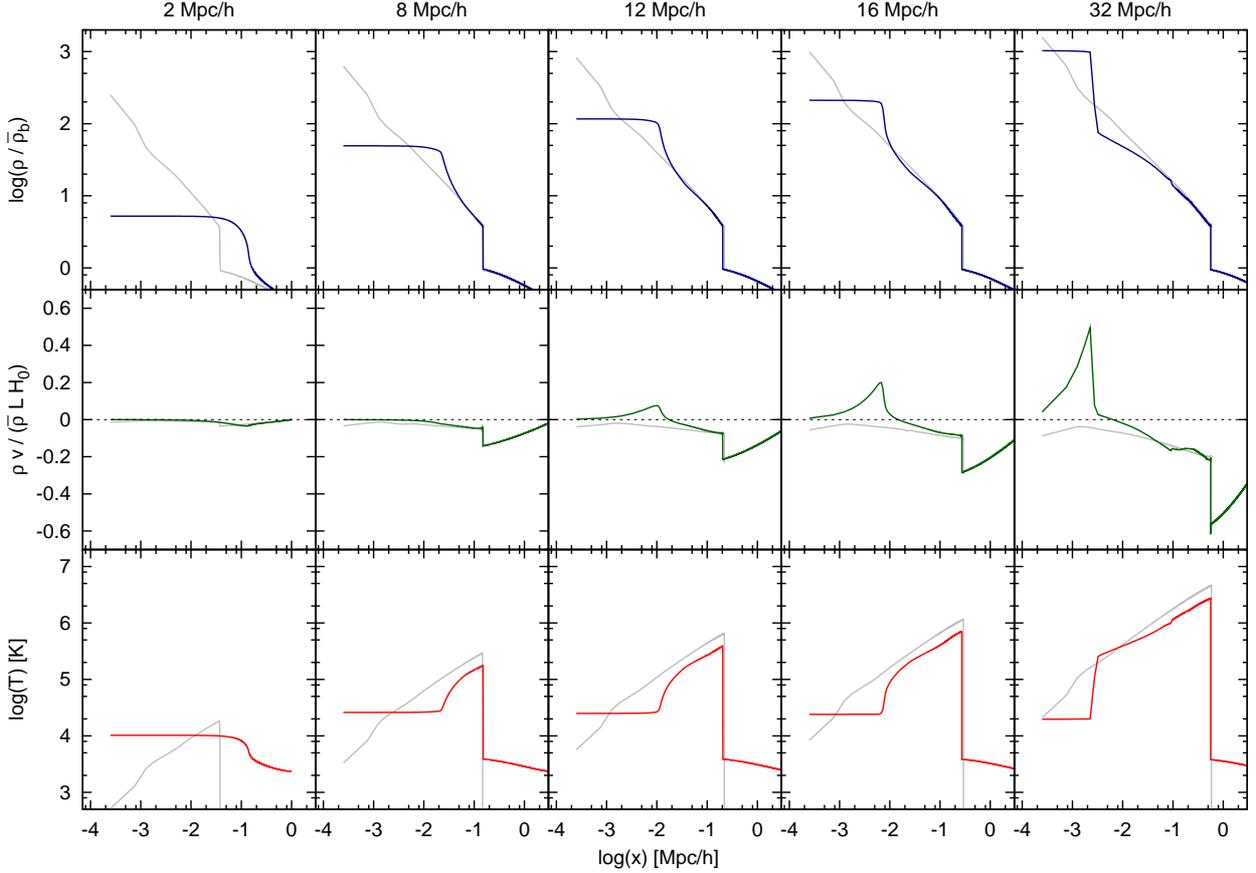}
\caption{Pancake formation including cooling and heating 
  for an initial perturbation amplitude of $A= 0.02$ and for a series of
  perturbation scales of $L=(2,8,12,16,32)$ Mpc/h comoving. Shown is the outcome of
  the simulations at $z= 0$ including cooling and heating (solid lines), and
  without dissipation (gray lines). The profiles are displayed using logarithmic
  coordinate axes.
  \emph{First row:} Density profiles; 
  \emph{Second row:} Density flux profiles; 
  \emph{Third row:} Temperature profiles.
  }
\label{fCool}
\end{figure*}

If cooling and heating are included into the consideration, an intrinsic physical
scale is introduced. Unlike before, the physical dimension cannot be eliminated
from the dynamical equations. Simulations using different perturbation scales
$L$ differ not only quantitatively but also qualitatively. As a consequence, the
constraint on the ratio between the spatial resolution $\Delta x$ and the local
Jeans length $\lambda_J$ as presented in \citet{Truelove97} has to be fulfilled:
\begin{equation}\label{eTruelove}
0.25 \ge \frac{\Delta x}{\lambda_J} 
= \Delta x \sqrt{\frac{G \, \rho_\mathrm{tot} \, \rho}{\pi \gamma p}}  \;.
\end{equation}
If this criterion is not fulfilled, the pressure is too weak to even out small
scale perturbations, which are caused by the finite numerical resolution. These
perturbations may then grow rapidly, and induce fragmentation for purely
numerical reasons. In one-dimensional simulations this violates
the spatial symmetry of the configuration. Without the inclusion of either a
(formal) heating source or an artificial pressure floor, catastrophic cooling
in the center will appear, i.e., the density increases while the pressure decreases. This
will always result in a Jeans length that violates Eq. (\ref{eTruelove}). For the
one-dimensional collapse, the heating due to the UV background is sufficient to
prevent such a cooling catastrophe.

In Fig. \ref{fCool} we present the outcome of our simulations including radiative
cooling and heating for different length scales $L$ of the initial perturbation.
For the computation of the chemical network, we assume IE, as described in Sect.
\ref{sDyn}. The influence of non-IE will be discussed in Sect. \ref{sChem}. Like
before, the initial amplitude is $A = 0.02$. With increasing $L$ the number of
grid points increases from $2000$ to $64000$, thus keeping a constant spatial resolution
of $0.5$ kpc. Displayed are the density, the density flux (instead of the
velocity, because it emphasizes the high-density region in the center), and the
related temperature profiles. We choose a logarithmic x-axis, thus focusing on
the center of the simulation box.

As an immediate effect of the heating due to the photoionizing UV background,
the configuration is heated up to temperatures of $T \approx 2 \times 10^4$ K during
the reionization at redshift $z\approx6$. This results in a pressure several
orders of magnitudes higher than in the non-radiative simulations. Therefore, the
adiabatic collapse before redshift ($z\approx1$) does not produce one single
peak, but an \emph{isothermal core} supported by pressure. The further evolution
now depends on the size of the perturbation length scale $L$.

For the smallest perturbation scale $L=2$ Mpc the speed of sound inside this core
remains always higher than the infall velocity, and therefore the shock cannot
form anymore. The whole profile, now sustained by the pressure of the gas
only, is more extended than in the non-radiative case. For $L>2$ Mpc the infall
velocity becomes higher than the sound velocity at some moment (this can be
obtained from the scaling considerations discussed above). Thus, a shock is able
to form. This shock is not generated in the center, but forms at the edges
of the pre-shock core. From there, it moves outward, like in the non-radiative
case. Additionally, a fan-like wave penetrates into the core, effectively
shrinking its size. The whole configuration can be separated into an inner
isothermal core, a shocked region of higher temperatures, and an outer region at
low density and low temperature. The size of the core is decreasing with
increasing length scale $L$. Outside of the core region, the results of the
simulations differ only slightly from the non-radiative case. Especially the
position of the outer shock appears to be unaffected. A special situation occurs
in the $L = 4$ Mpc simulation, where an effective outflow can be noticed, which
appears as a positive density flux \emph{outside} the core. This is caused by
the lower density inside the core compared to the runs with higher $L$, resulting
in longer cooling times, and thus a less effective dissipation of the energy
input by the further infall.

The influence of the perturbation scale on the properties of the core will be
further examined in Sect. \ref{sScaling}.

\subsection{Thermal conduction}

\begin{figure}
\centering
\resizebox{\hsize}{!}{\includegraphics{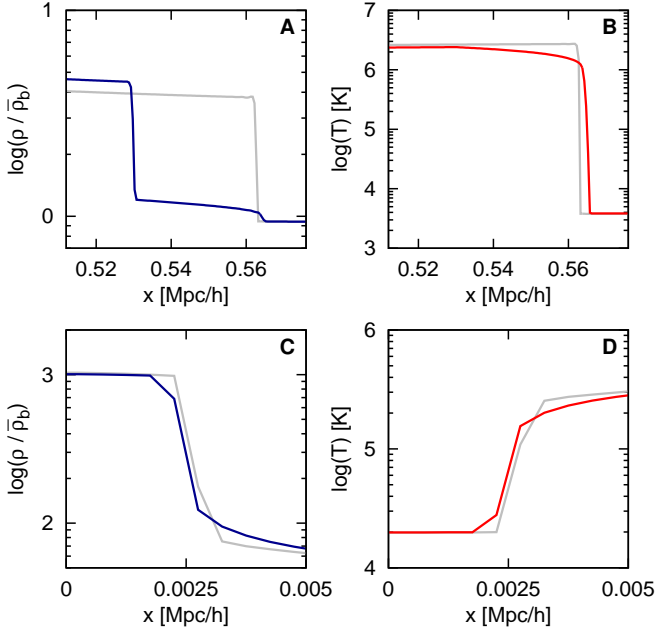}}
\caption{Influence of thermal conduction is shown at regions with suficciently steep
temperature gradients ($L = 32$ Mpc). The gray lines show the results of the
corresponding simulations without thermal conduction. \emph{Panel A and B:} The
density and temperature profiles at the outer shock fronts are shifted with
respect to the case without heat conduction. The high-temperature region is
extended outwards. The density is increased but over a lower volume (the density
shock is shifted inwards).
\emph{Panel C and D:} Heat conduction leads to a smaller core.} 
\label{fConduction}
\end{figure}

The inclusion of thermal conduction leaves an impact on the evolution of the
pancake only for perturbation scales of $L \ge 30$ Mpc. The used thermal
conduction coefficient $\kappa$ shows a steep temperature dependence of $\kappa
\propto T^{2.5}$. Moreover the temperature scales approximately like $T \propto
L^2$. This implies a relatively sharp transition between perturbation scales
where thermal conduction is effective or not. Furthermore, thermal conduction is
only important if steep temperature gradients occur. This is particularly
expected within the immediate neighborhood of the shock fronts or/and for the
temperature step at the core edges. Then it might occur that the thermal heat
conduction exceeds the cooling and the formed core is heated up (evaporation).
For this case, a rough order-of-magnitude estimate gives the condition under which
thermal conduction may overcome the cooling, i.e. $-\nabla \cdot\mathbf{j}>
\Lambda$. For an estimate, we assume an average cooling comparable with that by
recombination or collisional excitation. Then we get the relation
\begin{equation}\label{heatestimate}
n_e \lambda_\mathrm{T} < 10^{19} \,\left(\frac{T}{10^6 \mathrm{
K}}\right)^{7/4} \,
\mathrm{cm}^{-2} \;,
\end{equation}
where $n_e, \lambda_\mathrm{T}$ are the electron number density and the
characteristic scale for the temperature gradient, respectively. Owing to the considerable
temperature difference throughout the transition zone, one should take $T$ at the
at the core edge, whereas for $n_e$ one should take the value inside the core. At
temperature differences of about $10^6$ K and $n_e \approx 10^{-4}$ cm$^{-3}$ we
get a few kpc for the transition scale $\lambda_\mathrm{T}$.

In Fig. \ref{fConduction} we show the details of the density and temperature profiles
for one perturbation scale $L=32$ Mpc. The influence of $L$ will be further
discussed in Sect. \ref{sScaling}. The top panels show the region of the
confining shock. In the simulations including thermal conduction, the shocks in
density and temperature do not coincide anymore. Thermal energy from the shocked
region has been transported outward, heating the domain in front of the shock to
temperatures comparable to the shocked region. This energy is lost by the shocked
region, resulting in a lower pressure and causing a more confined density profile.
The now higher temperature in front of the shock results in a higher pressure
there, slowing down the infalling gas. This deceleration causes the small
increase in density adjacent to the shock.

Thermal conduction also affects the resulting core profile in the simulations
including heating and cooling (see Sect. \ref{sCool}). The inner part of the
pancake profile is shown in the bottom panels of Fig. \ref{fConduction}. Now, the
core is distinguished by a steep increase of the temperature at the core border
of approximately one order in magnitude. There, thermal conduction transports
energy toward the center of the profile. This additional energy raises the
pressure in the center, causing an outflow. This results in a smaller core size
with respect to the simulations without heat conduction. Contrary to the outer shock front, there is no displacement between the density and the temperature profiles. The core
shrinks as a whole.

\subsection{Chemical composition}\label{sChem}

\begin{figure}
\centering
\resizebox{\hsize}{!}{\includegraphics{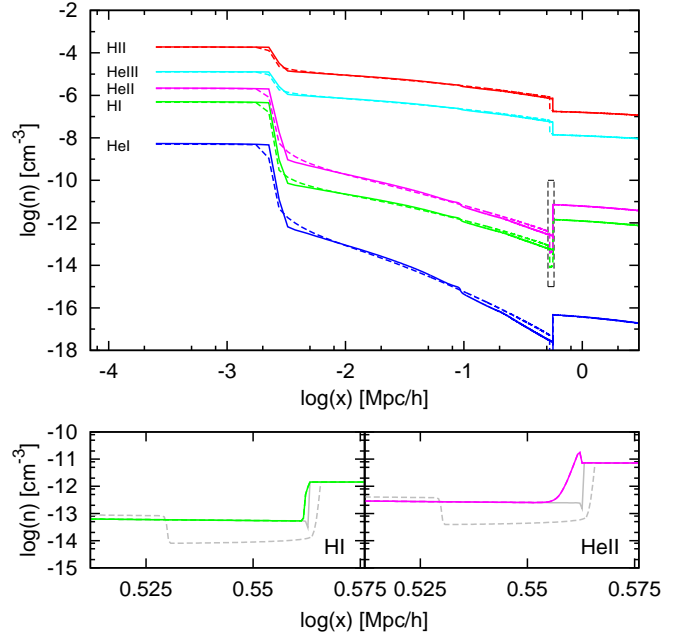}}
\caption{\emph{Top panel:} Chemical composition of a simulated pancake ($L = 32$
Mpc) including cooling and heating without thermal conduction (solid lines) and with
thermal conduction (dashed lines). The colors indicate
different species (lines from top to bottom): red: \ion{H}{II},
cyan: \ion{He}{III},
magenta: \ion{He}{II},
green: \ion{H}{I},
blue: \ion{He}{I};
\emph{Bottom Panel:} Chemical composition in the region of the shock (indicated by the box in the top panel) using
non-IE chemistry for \ion{H}{I} (left) and \ion{He}{II} (right). 
The corresponding profiles from the top panel are shown in gray for comparison.
[\emph{See the electronic edition of the Journal
for a color version of this figure.}] }
\label{fChem}
\end{figure}

Under quite general conditions, the characteristic time scales for chemical
processes, e.g., such as ionization and recombination, are much shorter than the
dynamical ones. Then, it is entirely justified to assume IE. The chemical
composition is entirely determined by the temperature and the number densities of
species engaged within the processes. In the top panel of Fig. \ref{fChem} we
show the number density profiles assuming IE for a perturbation scale of $L=32$
Mpc (this corresponds to the results shown in the fourth column in Fig.
\ref{fCool}). The gas is almost completely ionized, which leads to very low
abundances of \ion{H}{i}, \ion{He}{i}, and \ion{He}{ii}. The shapes of the
profiles resemble the density profile, except for a noticeable step at the
position of the shock caused by the steep decrease in temperature at this
position. In the simulations including thermal conduction, the behavior at the
position of the shock is even more complex: The number densities of the
\ion{H}{i}, \ion{He}{i}, and \ion{He}{ii} show a gap in the region adjacent to
the shock. Because of the offset between the temperature shock and the density shock
caused by thermal conduction (cp. with the preceding section), this region
combines a low density with a high temperature, which causes a higher degree of
ionization.

Under certain conditions the characteristic time scales become comparable and the
assumption of ionization equilibrium may become inappropriate. This might
happen at very small number densities. In this case, we have to follow the
detailed evolution of the chemical network using the full set of non-IE equations
(\ref{eN}).

In those simulations, omitting the assumption of IE, the results differ only
marginally from that ones of the IE-simulations. Effects on the hydrodynamic
evolution, coupled by the cooling/heating function to the chemical network,
cannot be detected at all. However, the chemical composition shows a slight deviation
with respect to the IE simulations, and this occurs only in the direct vicinity
of the shock. In the bottom panels of Fig. \ref{fChem} we show the number density
of \ion{H}{i} and \ion{He}{ii} around the shock using IE and non-IE. In the
non-IE simulation a very small region adjacent to the shock exists where for
\ion{He}{ii} the degree of ionization is lower than in the IE case. The
corresponding chemical timescale becomes comparable with the hydrodynamical
timescale. In combination with the motion of the shock, this produces the
observed behavior. Since, at the temperatures present in that region, the chemical
rates for the \ion{H}{i} are higher, the chemical time-scales are short, and a
similar feature is not observed. Owing to the discussed offset between temperature
shock and density shock, the chemical time scales at the shock region are even
larger, which results in a more extended region of delayed ionization. The same
behavior can be observed for \ion{He}{i}, only at much lower number densities.

Though the effects of non-IE for the primordial composition are only marginal,
the influence of the particular conditions at the shock position has been
demonstrated. Thus for a medium containing some fraction of heavy elements which
have very low densities for usual abundances, the non-IE must be taken into
consideration.

If omitting the assumption of uniform temperature for all fluid components
(electrons, ions) the effects on non-IE maybe much larger \citep{Teyssier98}.

\subsection{Influence of small-scale perturbations}\label{sGef}

\begin{figure}
\centering
\resizebox{\hsize}{!}{\includegraphics{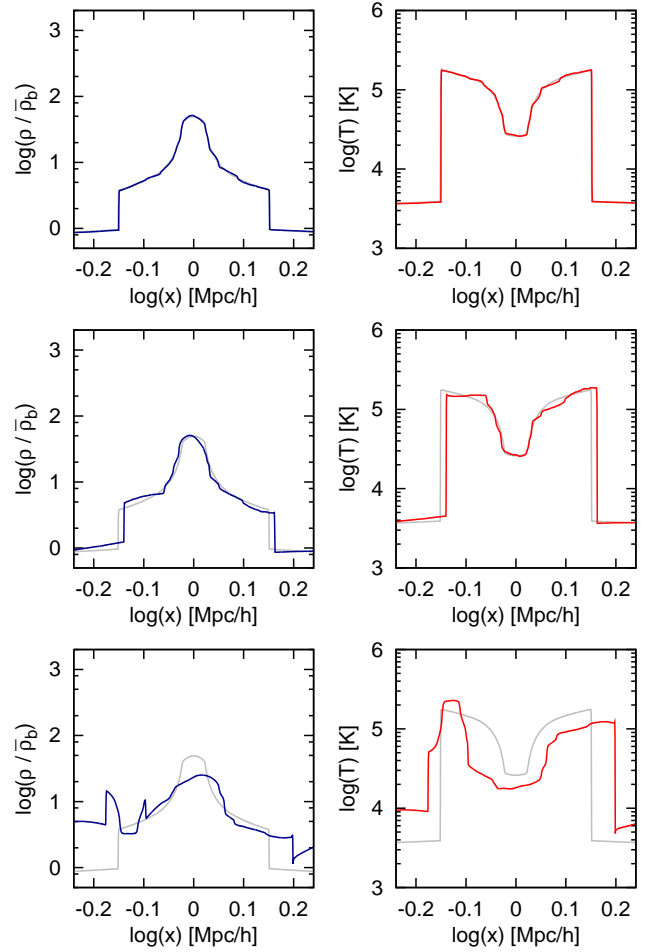}}
\caption{Pancake formation if Gaussian random perturbations are taken into account for the initial density field according to the cosmological perturbation power spectrum. The resulting density (\emph{left}) and temperature (\emph{right}) profiles are shown at \mbox{$z=0$} for an initial perturbation at $L=8$ Mpc. The included upper scale length for the Gaussian random perturbations as fraction of $L$ decreases from top to bottom: 
\emph{First row:} 1/8 $L$;
\emph{Second row:} 1/4 $L$;
\emph{Third row:} 1/2 $L$ . 
The obtained profiles are compared with the single-mode consideration given in \mbox{Fig. \ref{fCool}}. The latter is shown in gray. 
}
\label{fGef}
\end{figure}

The initial conditions for cosmological simulations of large-scale structure formation are given by a spectrum of perturbations. To take this into account, we add Gaussian random perturbations according to the cosmological power spectrum to the particular perturbation given by Eq. (\ref{ePancake}). Then, in a given spatial region of size $L$, a pronounced pancake structure will only form if the considered initial perturbation amplitude dominates over the neighboring perturbation amplitudes at comparable scale size. Thus, we subsequently include all perturbation modes up to a scale size of $(1/8,1/4,1/2)\times L$. 

In Fig. \ref{fGef} we show the density and temperature profiles of simulations including the small scale perturbations and cooling and heating at $z=0$. Clearly, the perturbation modes at scales comparable with $L$ have the most impact on the final profiles at $z=0$. In this case most of the power from perturbation modes at neighboring scales will be added. In particular, this leads to an enhancement of the core density. The modes smaller than the actual Jeans length are erased by the heating due to the UV background. However, the magnitudes of the temperatures and their coarse profiles in the post-shock region are on the same order and the density profiles are nearly preserved. 

We conclude that it is sufficient to consider only the collapse of a single (large enough) mode in order to gain \emph{coarse} information on the thermal and chemical state of the structures of interest. In addition, the conservation of the system's symmetry serves as a probe for the quality of the numerical treatment. The non-radiative simulations reproduce the earlier found analytic results with high accuracy \citep{Shandarin89}.

\section{Scaling relations}\label{sScaling}

\begin{figure}
\centering
\resizebox{\hsize}{!}{\includegraphics{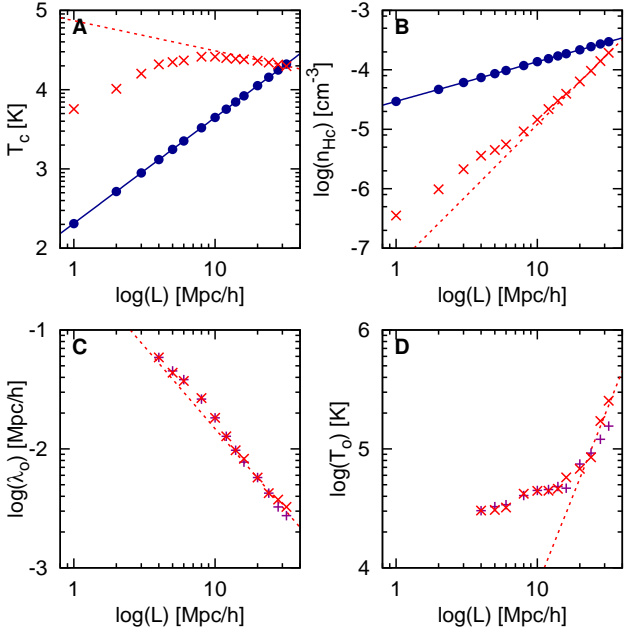}}
\caption{One-dimensional collapse: Dependence of various quantities on the
perturbation length scale $L$ for simulations without (${\color{blue}\bullet}$),
with (${\color{red}\times}$) cooling and heating, and with cooling and heating
and thermal conduction (${\color{violet}+}$). 
\emph{Panel A}: Central temperature (the dotted line represents $T_c\propto L^{-0.38}$; 
\emph{Panel B}: Central hydrogen number density (the dotted line represents $n_\ion{H}{} \propto L^{2.38}$); 
\emph{Panel C}: Core size (the dotted line represents $\lambda_{\circ}\propto L^{-1.38}$);
\emph{Panel D}: Temperature at the core boundary (the dotted line represents
$T_{\circ} \propto L^{3}$). In the upper panels the data points with thermal
conduction are identical to those without and are therefore omitted.}
\label{fScaling}
\end{figure}

\begin{figure}
\centering
\resizebox{\hsize}{!}{\includegraphics{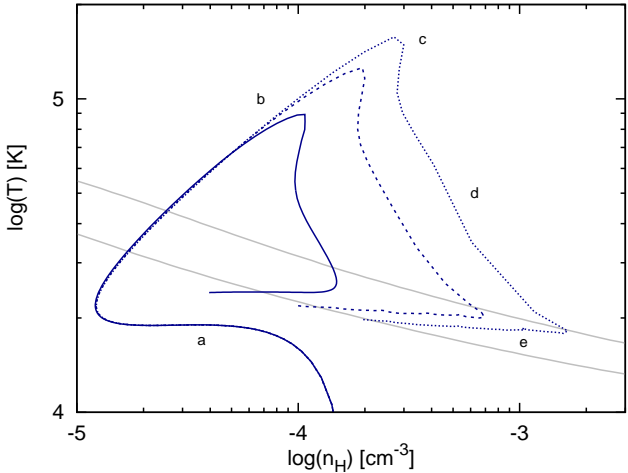}}
\caption{Phase space diagram, i.e., the dependence temperature versus
density, is shown for the quantities at the center of the pancake. Shown are curves for $L=(16,24,32)$ Mpc using \emph{solid}, \emph{dashed}, and \emph{dotted} lines, respectively. The upper gray line shows equilibrium temperature at redshift of $z=0.7$ (time of shock formation) and the lower gray line at $z=0$. After a period of linear growth (a), the perturbation decouples from the cosmic expansion and contracts adiabatically (b) until shock formation (at the maximum of the temperature). Then, after nearly isochoric (c) and isobaric (d) evolution stages the central gas arrives at thermal equilibrium (e). 
}
\label{fPhase}
\end{figure}

\begin{figure}
\centering
\resizebox{\hsize}{!}{\includegraphics{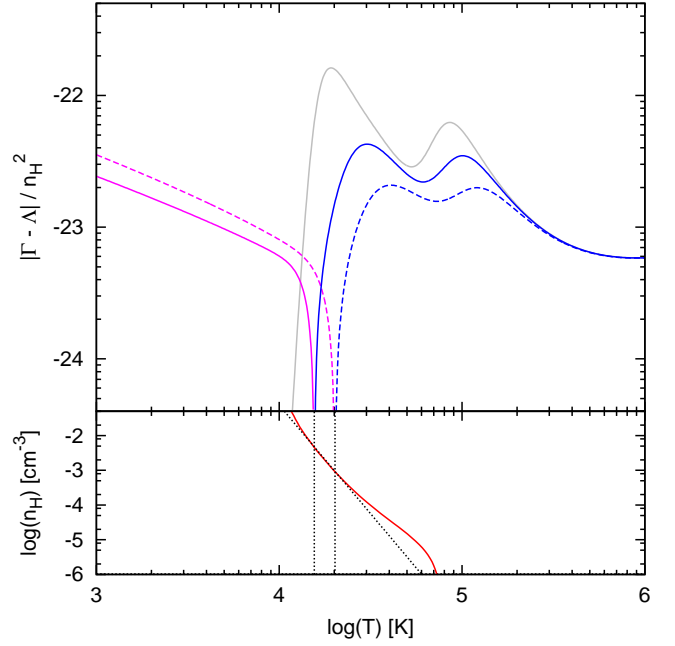}}
\caption{\emph{Top panel:} Absolute value of the cooling/heating function
$|\Gamma - \Lambda|$ for a hydrogen number density of $n_\mathsc{H} = 4.6 \times 10^{-3}$ cm$^{-3}$
corresponding to an overdensity of $\rho/\bar{\rho}_b = 5000$ at $z=0.7$ (solid
line) and for a density of $n_\mathsc{H} = 9.2 \times 10^{-4}$ cm$^{-3}$ corresponding to
an overdensity of $\rho/\bar{\rho}_b = 1000$ (dashed line). The cooling function
without heating is given in gray for comparison. \emph{Bottom panel:} Equilibrium
temperature vs. density. In the range delimited by the two densities used in the
top panel (emphasized by the the dashed lines) the relation can be approximated
by a power law $T_e \propto n_{\mathsc{H}}^{-0.16}$.}
\label{fCooling}
\end{figure}

In the preceding section it was shown that the length scale of the perturbation $L$ 
is the determining parameter for the
evolution of the pancake characteristics (temperature and density profiles). In Fig.
\ref{fScaling} we present the $L$-dependence of the final values of the central
temperature $T_c$, the central hydrogen number density $n_{\mathsc{H}c}$, the radius of the
isothermal core $\lambda_\circ$, and the temperature at its edge $T_\circ$.

For the non-radiative simulations, $T_c$ shows the expected behavior: The
temperature scales $\propto L^2$, and thus $T_c$, as well. The density does
not depend on $L$, its profile is uniquely $\propto r^{-2/3}$
\citep[see][]{Shandarin89}. However, because we increase the number of grid points
in order to keep the spatial resolution fixed, larger simulations resolve the
central peak better.  In the result, we get an apparent dependence of the central
density, i.e., of the density at the innermost grid point, $\propto L^{2/3}$.
However, the latter reflects only the degree of resolution.

If including cooling and heating processes, the above relations are no longer
valid. The central temperature stays roughly constant at about $2\times10^4$ K, 
weakly decreasing at higher $L$ proportional to $L^{-0.38}$. The central hydrogen number 
density shows a strong dependence on $L$ approaching the asymptote $n_{\mathsc{H}c} \propto
L^{2.38}$. At the edge of the isothermal core, the temperature strongly increases
outwards, but then flattens again. We identify the core radius $\lambda_\circ$ as
the distance from the center to the inflexion point of the temperature profile.
Then the core radius shows a dependence on the scale length proportional to
$L^{-1.38}$.

The above mentioned scaling relations for $T_c$, $n_{\mathsc{H}c}$, and $\lambda_\circ$
can be explained using simple thermodynamical arguments. Actually, the size of
the core and its density is fully determined by conditions of hydrostatic
equilibrium. In this case Euler's equations yield for the central pressure $p_c
\sim n_{\mathsc{H}c} \phi$ and Poisson's equation yields $\phi \sim n_{\mathsc{H}c}
\lambda_\circ^2$. Combining those two estimates we obtain
\begin{equation}\label{eCore}
  p_c \sim \lambda_\circ^2 \, n_{\mathsc{H}c}^2
\end{equation}
Without heating the actual pressure of the cooling gas is not able to withstand
the gravity forces, and no distinguishable core is forming in these simulations.
Including heating, the pressure is raised by several orders of magnitude during
reionization at \mbox{$z \sim 6$}.

In Fig. \ref{fPhase} we show the path with respect to the central point of the configuration
in a phase space diagram, where the temperature $T_c$ is plotted against the
hydrogen number density $n_{\mathsc{H}c}$ (these are physical, not
super-comoving quantities). The curve enters the shown domain at
the center of the bottom of the plot, when reionization heats the gas to $\approx
2\times10^4$ K. From there, the density decreases at an almost constant
temperature according to the linear perturbation growth and cosmological expansion. 
Then the gas dynamics decouples from the Hubble flow and follows the
adiabatic pressure-density relation $p_c \propto n_{\mathsc{H}c}^{\gamma}$.
The shock forms at the maximum of $p_c$ and $T_c$. During
the collapse, the energy budget of the central region is dominated by the infalling matter
and not by cooling and heating. Therefore, the scaling relation for the pressure
derived for the non-radiative simulations $p \propto L^2$ also holds here.
Using this relation (and $\gamma = 5/3$) we obtain scaling relations for the
central density and the central temperature at the time of shock formation
(denoted by the index $s$):
\begin{eqnarray}
n_{\mathsc{H}s} &\propto& p_s^{1/\gamma} \propto L^{2/\gamma} = L^{1.2}\\ 
T_s &\propto& p_s / n_{\mathsc{H}s} \propto L^{2 - 2/\gamma} = L^{\frac{2 \left(1 -
/\gamma\right)}{\gamma}} = L^{0.8} \;.
\end{eqnarray}
The formation of the two shocks moving outwards changes the situation
significantly. The energy supply through infall vanishes, and the subsequent
evolution is dominated by cooling and heating. Nevertheless the pressure inside
the shocked region, determined by the potential, stays roughly constant, like in
the case without cooling and heating. Therefore the relation $p_c \propto L^2$
and Eq. (\ref{eCore}) remains valid. The efficient radiative cooling decreases
the temperature toward the equilibrium temperature $T_e$ of the cooling/heating
function, which is the temperature where the contributions of the cooling processes
and the UV background heating are canceling each other. The gas with a higher
temperature cools down toward $T_e$, while colder gas gets heated. In case of
ionizational equilibrium this temperature is a function of the density only. In
the top panel of Fig. \ref{fCooling} we show the absolute value of the normalized
cooling/heating function $|\Gamma - \Lambda|/n^2_{\ion{H}{}}$ for number
densities corresponding to overdensities of 1000 and 5000 at redshift $z=0.7$,
which is the approximate time of shock formation. The equilibrium temperature is
given by the zero value of $|\Gamma - \Lambda|/n_{\mathsc{H}}^2$ (the values in
Fig. \ref{fCooling} are absolute values). A higher density shifts $T_e$ 
toward lower temperatures. Computing $|\Gamma -
\Lambda|/n_{\mathsc{H}}^2$ for several densities and tracking the minimum we
obtain a relation between $n_{\mathsc{H}}$ and $T_e$. This is shown in the bottom
panel of Fig. \ref{fCooling}. This relation is also shown in Fig. \ref{fPhase}
(solid gray lines), for both $z=0.7$ and $z=0$. Evidently, when reaching $T_e$,
the central state remains there, evolving further only due to the cosmological
expansion. In the interval of overdensities of $1000$-$5000$ (at $z=0.7$), 
the relation can be approximated by a power law $T_e\propto n_{\mathsc{H}}^{-0.16}$ 
(indicated in Fig. \ref{fCooling}). At lower densities the behavior deviates only weakly from that
relation. Using this approximation and $p_c \propto n_{\mathsc{H}c} T_c \propto n_{\mathsc{H}c}^2
\lambda_\circ^2 \propto L^2$ we obtain
\begin{eqnarray}
n_{\mathsc{H}c} &\propto& p_c / T_c \propto p_c / n_{\mathsc{H}c}^{-0.16} 
\Rightarrow n_{\mathsc{H}c} \propto L^{2.38} \\ 
T_c &\propto& n_{\mathsc{H}c}^{-0.16} \propto L^{-0.38} \\
\lambda_\circ &\propto& L / n_{\mathsc{H}c} \propto L^{-1.38} \;.
\end{eqnarray}
Although we have only used the approximated relation for the equilibrium
temperature, the above consideration leads us to expressions that fully agree
with the results shown in Fig. \ref{fScaling}. 
This means that the system eventually approaches a quasi-equilibrium state for sufficiently large scales, at least.

The core size decreases with increasing scale length $L$ according to
$\lambda_{\circ}\propto L^{-1.38}$. As can be seen in Figs. \ref{fConduction} and
\ref{fScaling} the heat conduction even amplifies this tendency. Thus it might be
expected that for some scale length $L$ an evaporation of the core will happen.
Obviously, that will be the case if the core according to the cooling/heating
processes becomes of comparable size as the above introduced heat transition
scale $\lambda_T$, i.e. $\lambda_T \ge \lambda_{\circ}$. Using the expression
(\ref{heatestimate}) and the derived scaling relations normalized to the
corresponding numerical values at $L=16$ Mpc we obtain for the scale length at
which the core is significantly affected by thermal conduction
\begin{equation}
L \approx 30 \textrm{ Mpc} \;.
\end{equation}
At larger scales even evaporation of the core becomes possible. 
This also principally agrees with the results obtained by \citet{Bond84}.

\section{Summary and conclusions}

The aim of the present paper is to investigate the detailed physics and
thermodynamics of a matter distribution which reflects the basic properties of
structures of the WHIM. From cosmological simulations we know that the WHIM is
distributed in sheets and filaments. For the description
of a matter distribution at low density the resolution is a key issue. Therefore,
we initially concentrated on the description of the one-dimensional
collapse toward a gaseous sheet at extremely high resolution. Although
neglecting any interaction between the perturbations on various scales we
started with perturbation parameters consistent with the cosmological density field.
Above an initial perturbation scale of $\approx 2$ Mpc the collapsing gas shocks
and further thermodynamics are mainly determined by the cooling and heating
processes. In case of the one-dimensional collapse the heating by the
photoionizing UV background is sufficient to keep the whole structure in a
quasi-equilibrium state. Then the final density and temperature distribution
depends on the initial perturbation length scale $L$ only, i.e. all quantities
characterizing the quasi-equilibrium state may be roughly described as function of
that length scale $L$. If keeping the resolution high enough and fixed, and
increasing the initial length scale $L$ then this goes beyond the numerical
capabilities. The obtained scaling relations can be used thought to extrapolate
the results towards higher $L$.

The one-dimensional density and temperature profiles are characterized by the
existence of a cold and dense core region which is at thermal equilibrium at a
temperature of $T_c\approx 2 \times 10^4$ K. This core is built even before shock
formation, and its properties are given by the interplay between radiative
cooling and the energetic input from the UV background. 

In this respect the question about a possible shielding of the UV flux is essential. However, the estimates of the optical depth $\tau_\mathrm{ph}$ with respect to photoionization by the UV flux show for all possible cases $\tau_\mathrm{ph} \ll 1$. The situation may be different for the three-dimensional case when higher densities are reached.

The determining variable
is again the perturbation scale $L$: larger collapsing scales lead to a spatially
narrower cold region. The size of the core region decreases even more if thermal
conductivity becomes efficient. For large enough scales $L$ the temperature
gradients at the transition from the cold core towards the shock heated gas are
large. In the result, thermal conductivity leads to a partial "evaporation" of
the core. Using the derived scaling relations with respect to the parameter $L$ we
can estimate the approximate collapse scale when the core will be entirely
evaporated. In the result, we get a range of scales $L$ between 2 and 30 Mpc, for
which a cold and shock-confined core can exist. 

The cold gas in the core region of the WHIM structures must contain a
relatively large fraction of neutral hydrogen and should be detectable as
Ly$\alpha$ absorption lines in the light of more distant sources. If the core
size is estimated to be on the order of $\approx$ 10 kpc and the overdensity is
in the range $100 < \delta < 1000$, then Ly$\alpha$ absorption lines could reach
column densities of about $N_{\ion{H}{I}} \approx 10^{14}$ to $10^{15}$
cm$^{-2}$. At once, a contribution in absorption or emission by the heavy
element contamination in the WHIM should be detected. A likely detection of such
a simultaneous event was reported recently by \citet{Nicastro09}, using
combined X-ray and optical observations towards the Seyfert 1 Galaxy PKS
0558-504. Of course, the particular observation depends on the actual geometry
and orientation of the filament and the line of sight. The probability for direct
observations of the cold core region is rather low.

\begin{figure}
\centering
\resizebox{\hsize}{!}{\includegraphics{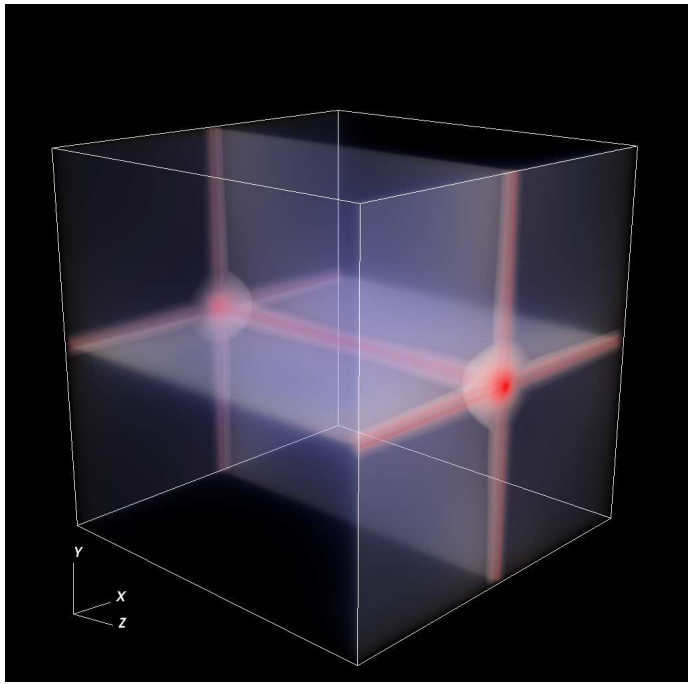}}
\caption{Rendering of the density field at $z = 0$ for three-dimensional
simulation using $L = 4$ Mpc. Overdensities in red, underdensities in blue
[\emph{See the electronic edition of the Journal for a color version of this figure.}]
}
\label{fFilament}
\end{figure}

High-resolution hydrodynamical simulations on galaxy formation \citep{Ocvirk08,Keres09a,Keres09b,Agertz09,Ceverino09} indicate another very important aspect. At higher redshifts $z = 2 - 3$, relatively cold gas is flowing along the filaments toward the connected galactic halos. The cold gas streams may even reach inner galactic regions, which are closer to the center than the virial shock region. These cold gas streams may supply galaxies with fuel for star formation or may influence the the galactic star formation, at least. Several studies discuss in which way the existence of these cold streams is related to the galaxy mass. \citet{Keres05} find that the cold gas accretion is only important for total galactic masses $\le 10^{11.4} M_{\sun}$. Related estimates are given in \citet{Dekel09}. If the WHIM gas is located mainly within a network of large shock-confined filaments at $z \approx 0$, then even recently a certain gas flow could be expected towards the corresponding knots of the network. Those knots could be matter concentrations of galactic or even of galaxy cluster size.

If extending our above considerations in three dimensions, multi-streaming occurs, leading to the formation of a well-defined filament and a halo, cp. Fig. \ref{fFilament}. Though the shape of the gas distribution is extremely idealized, it is expected to exhibit the correct density and temperature profiles depending on the initial scale length, again. The large-scale gas velocities are determined by the gravitational potential. In the three-dimensional case, the velocities toward the forming filament and even more toward the halo are becoming high enough to generate shocks. Instead of a core, a relatively cold stream forms along the three-dimensional filament, propagating deep into the halo. The temperature and density profiles for the three-dimensional case are much more complex even for the highly idealized geometry. We are going to consider this in a forthcoming paper (Klar \& M\"ucket, in preparation). However, the existence of a natural regulation mechanism for the size and existence of a cold core for the one-dimensional case might indicate similar restrictions for the cold stream in three dimensions.

\begin{acknowledgements}
JSK acknowledges A. Partl, T. Doumler, S. Knollmann, A. Booley and R. Teyssier
for help and useful discussions. JSK was supported by the Deutsche
Forschungsgemeinschaft under the project MU 1020/6-4.s and by the German Ministry
for Education and Research (BMBF) under grant FKZ 05 AC7BAA.
\end{acknowledgements}

\bibliography{paper}{}	
\bibliographystyle{aa}

\appendix
\section{chemical rates and a model for the UV background}
\label{aUV}

% chem table
\begin{table*}
\caption{Used rates for chemical evolution, cooling, and heating. Rates of
collisional processes are taken from \citet[Tables 1 and 2]{Katz96};
Photoionization and photoheating rates are taken from \citet[Sect. 2.1]{Black81}.
$T_\alpha$ denotes $T / 10^\alpha$ K.}
\label{tChemicalRates}
\centering
\begin{tabular}{lllll}
\hline\hline
Process & Type & Species & Coefficient & Value \\ 
\hline
standard recombination & chemical & \ion{H}{ii} & 
$\alpha_{\ion{H}{ii}}$ &
$8.40 \times 10^{-11} \, T^{-0.5} \, T_3^{-0.2} \, ( 1 + T_6^{0.7} )^{-1}$
\\
&& \ion{He}{ii} & 
$\alpha_{\ion{He}{ii}}$ & 
$1.50 \times 10^{-10} \,  T^{-0.6353}$
\\
&& \ion{He}{iii} & 
$\alpha_{\ion{He}{iii}}$ &
$3.36 \times 10^{-10}  \, T^{-0.5} \, T_3^{-0.2} \, ( 1 + T_6^{0.7} )^{-1}$
\\
& cooling & \ion{H}{ii} & 
$\eta_{\ion{H}{ii}}$ & 
$8.70 \times 10^{-27} \, T^{0.5} \, T_3^{-0.2} \, ( 1 + T_6^{0.7} )^{-1}$
\\ 
&& \ion{He}{i} & 
$\eta_{\ion{He}{i}}$ &
$1.55 \times 10^{-26} \, T^{0.3647} $
\\
&& \ion{He}{ii} & 
$\eta_{\ion{He}{ii}}$ &
$3.48 \times 10^{-26} \, T^{0.5} \, T_3^{-0.2} \, ( 1 + T_6^{0.7} )^{-1}$
\\
\hline
dielectric recombination & chemical & \ion{He}{ii} & 
$\xi_{\ion{He}{ii}}$ & 
$1.90 \times 10^{-3\hphantom{1}}
\, T^{-1.5} \, \exp( -470000 / T ) \,
( 1 + 0.3 \exp( -94000 / T ) )$
\\
& cooling & \ion{He}{ii} &
$\omega_{\ion{He}{ii}}$ & 
$1.24 \times 10^{-13} \, T^{-1.5} \, \exp( -470000 / T ) \,
( 1 + 0.3 \exp( -94000 / T ) )$
\\
\hline
collisional ionization & chemical & \ion{H}{i} & 
$\beta_{\ion{H}{ii}}$ & 
$5.85 \times 10^{-11} \, T^{0.5} \, ( 1 + T_5^{0.5} )^{-1}
\exp( -157809.1 / T )$
\\ 
&& \ion{He}{i} & 
$\beta_{\ion{He}{ii}}$ &
$2.38 \times 10^{-11} \, T^{0.5} \, ( 1 + T_5^{0.5} )^{-1}
\exp( -286335.4  / T )$
\\
&& \ion{He}{ii} & 
$\beta_{\ion{He}{iii}}$ &
$5.68 \times 10^{-12} \, T^{0.5} \, ( 1 + T_5^{0.5} )^{-1}
\exp( -631515.0 / T )$
\\
& cooling & \ion{H}{i} & 
$\zeta_{\ion{H}{i}}$ &
$1.27 \times 10^{-21} \, T^{0.5} \, ( 1 + T_5^{0.5} )^{-1} 
\exp( -157809.1 / T)$
\\
&& \ion{He}{ii} & 
$\zeta_{\ion{He}{ii}}$ & 
$1.27 \times 10^{-21} \, T^{0.5} \, ( 1 + T_5^{0.5} )^{-1}
\exp( -157809.1 / T )$
\\
&& \ion{He}{iii} & 
$\zeta_{\ion{He}{iii}}$ &
$4.95 \times 10^{-22} \, T^{0.5} \, ( 1 + T_5^{0.5} )^{-1}
\exp( -631515.0 / T )$
\\
\hline
photoionization & chemical &  \ion{H}{i} & 
$\gamma_{\ion{H}{ii}}$ &
$2.54 \times 10^{8} \, j_0$
\\ 
&& \ion{He}{i} & 
$\gamma_{\ion{He}{ii}}$ &
$2.49 \times 10^{8} \, j_0$ 
 \\
&& \ion{He}{ii} & 
$\gamma_{\ion{He}{iii}}$ &
$1.60  \times 10^{7} \, j_0$ 
\\
& heating & \ion{H}{i} & 
$\varepsilon_{\ion{H}{i}}$ &
$7.75 \times 10^{-12} \, \gamma_{\ion{H}{i}}$
\\ 
&& \ion{He}{i} & 
$\varepsilon_{\ion{He}{i}}$ &
$2.19 \times 10^{-11} \, \gamma_{\ion{He}{i}}$ 
\\
&& \ion{He}{ii} & 
$\varepsilon_{\ion{He}{ii}}$ &
$3.10 \times 10^{-11} \, \gamma_{\ion{He}{ii}}$ 
\\
\hline
collisional excitation & cooling & \ion{H}{i} &
$\psi_{\ion{H}{i}}$ &
$7.50 \times 10^{-19} \, ( 1 + T_5^{0.5} )^{-1}
\exp( -118348.0 / T )$
\\
&& \ion{He}{ii} &
$\psi_{\ion{He}{ii}}$ &
$5.54 \times 10^{-17} \, T^{-0.397} \, ( 1 + T_5^{0.5} )^{-1}
\exp( -473638.0 / T )$
\\
\hline
bremsstrahlung & cooling & all ions & 
$\theta$ &
$1.42 \times 10^{-27} \, g_{ff} \, T^{0.5} \quad$ 
with the gaunt-factor $g_{ff} = 1.5$
\\
\hline
\end{tabular}
\end{table*}

The coefficients used to compute the evolution of the chemical network and the
cooling and heating function are shown in Table \ref{tChemicalRates}. The values
of the rates of the collisional processes are taken from \citet{Katz96}, while
the photoionization and photoheating rates are taken from \citet{Black81}. In
terms of these rates the chemical source term writes:
\begin{eqnarray}
\Xi_\ion{H}{i} &=& 
\alpha_{\ion{H}{II}} \, n_{\ion{H}{II}} \, n_e
- \left( \beta_{\ion{H}{I}} \, n_e + \gamma_{\ion{H}{I}} \right) \, n_{\ion{H}{I}} \\
\Xi_\ion{H}{ii} &=& 
\left( \beta_{\ion{H}{I}} \, n_e + \gamma_{\ion{H}{I}} \right) \, n_{\ion{H}{I}}
- \alpha_{\ion{H}{II}} \, n_{\ion{H}{II}} \, n_e \\
\Xi_\ion{He}{i} &=& 
\left(\alpha_{\ion{He}{II}}+\xi_{\ion{He}{II}} \right) n_{\ion{He}{II}} \, n_e
\nonumber \\ 
&& - \left( \beta_{\ion{He}{I}} \, n_e - \gamma_{\ion{He}{I}} \right)
\, n_{\ion{He}{I}} \\
\Xi_\ion{He}{ii} &=&
\alpha_{\ion{He}{III}} \, n_{\ion{He}{III}} \, n_e 
\left( \beta_{\ion{He}{I}} \, n_e  - \gamma_{\ion{He}{I}} \right) \,n_{\ion{He}{I}} 
\nonumber\\ 
&& - \left( \alpha_{\ion{He}{II}}+\xi_{\ion{He}{II}} \right) 
n_{\ion{He}{II}} \, n_e \nonumber \\
&& - \left( \beta_{\ion{He}{II}} \, n_e +
\gamma_{\ion{He}{II}} \right) \, n_{\ion{He}{II}} \\
\Xi_\ion{He}{iii} &=&
\left(\beta_{\ion{He}{II}} \, n_e + \gamma_{\ion{He}{II}} \right) n_{\ion{He}{II}}
- \alpha_{\ion{He}{III}} n_{\ion{He}{III}} \, n_e \;.
\end{eqnarray}
The cooling function $\Lambda$ is the sum of contributions from the
collisional processes discussed above as well as collisional excitation of 
\ion{H}{I} and \ion{He}{II} and bremsstrahlung, while the heating function 
$\Gamma$ is the sum of the heating rates corresponding to the
photoionization: 
\begin{eqnarray}
\Lambda &=& \zeta_{\ion{H}{i}} \, n_{\ion{H}{i}} \, n_e 
+ \zeta_{\ion{He}{i}} \, n_{\ion{He}{i}} \, n_e
+ \zeta_{\ion{He}{ii}} \, n_{\ion{He}{ii}} \, n_e \nonumber\\
&& 
+ \; \eta_{\ion{H}{ii}} \, n_{\ion{H}{ii}} \, n_e
+ \eta_{\ion{He}{ii}} \, n_{\ion{He}{ii}} \, n_e
+ \eta_{\ion{He}{iii}} \, n_{\ion{He}{iii}} \, n_e \nonumber\\
&&
+ \; \omega_{\ion{H}{ii}} \, n_{\ion{H}{ii}} \, n_e 
+ \psi_{\ion{H}{i}} \, n_{\ion{H}{i}} \, n_e 
+ \psi_{\ion{He}{ii}} \, n_{\ion{He}{ii}} \, n_e \nonumber\\
&&
+ \; \theta \left(n_{\ion{H}{ii}} + n_{\ion{He}{ii}} + 4 \, n_{\ion{He}{iii}}
\right) \, n_e
\\
\Gamma &=& \varepsilon_{\ion{H}{i}} \, n_{\ion{H}{i}} 
+ \varepsilon_{\ion{He}{i}} \, n_{\ion{He}{i}} 
+ \varepsilon_{\ion{He}{ii}}\, n_{\ion{He}{ii}} \;.
\end{eqnarray}

For the computation of photoionization as well as photoheating the flux of the
UV background is needed. We use a simplified model for its redshift dependence
which resembles the current view in the literature.
\citep{Gnedin00,HaardtMadau01,Bianchi01}:
\begin{equation}
j_0 = 10^{-21} \, \mathrm{erg} \, \mathrm{s}^{-1} \mathrm{cm}^{-2}
	\times \left\{\begin{array}{ll} 
           10^{-5} & \textrm{if } z > 8   \\
           0.5 \times 10^{0.35 \left(6 - z\right) } & \textrm{if } 8\le z <6 \\
           10^{0.1 \left(3 - z\right)} & \textrm{if } 6 \le z < 3  \\
           1 & \textrm{if } 3 \le z < 1  \\
           0.1 \times 10^z & \textrm{if } z \le 1  \\
           \end{array} \right. \;.
\end{equation}
A spectrum inversely proportional to the frequency is assumed.

\section{The \texttt{evora} code}
\label{aCode}

Most of the techniques used in our code are common in cosmological simulation
codes. We will therefore concentrate on specific features of \texttt{evora}. To
preserve readability we will give one dimensional descriptions in $x$ direction.
The generalization to three dimensions is straightforward. We will use $x^\prime
= x(t +\Delta t)$ for a quantity $x$ after the time step $\Delta t$ and $x =
x(t)$ at its beginning.

\subsection{Time stepping}

The length of the time step $\Delta t$ is given by the minimum of
three different constraints. The first is the
Courant-Friedrich-Levy condition implied by the the hydrodynamic solver:
\begin{equation}
\Delta t_{\mathrm{cfl}} = \frac{\Delta x}{|u| + a} \;.
\end{equation}
where $a$ is the speed of sound. Moreover, the
cosmological expansion during one single time step is limited by
\begin{equation}
\Delta t_{a} = \frac{a}{\dot{a}} \;.
\end{equation}
The last constraint on the time step is associated with thermal conduction.
Here, the maximal length of the time step can be computed by estimating the 
fraction between the thermal energy and and its change due to the thermal 
conduction:  
\begin{eqnarray}
\Delta t_{\mathrm{tc}} = \frac{E_{\mathrm{th}}}{\dot{E}_{\mathrm{th}}} 
\approx \frac{\rho T}{\kappa T / \Delta x^2}
= \frac{\rho \Delta x^2}{\kappa} \;.
\end{eqnarray}
Furthermore, each of these time steps is further restricted by a heuristic
factor $0 < C < 1$. We use $C_{cfl} = 0.5$ for the CFL-constaint, $C_{a} =
0.01$ for the cosmological constraint, and $C_{tc} = 0.9$ for the thermal
conduction constraint. The time steps are computed in every cell, and the
minimal value is used to compute the global time step.

The cooling timescale (which is also the chemical timescale) can be much shorter 
than timescales given above. Therefore, it is
not used as a constraint on the global time step. Instead we
use \emph{subcycling}: The time evolution of the number densities and the
thermal energy density is computed using several shorter time steps. 
This is done locally in each cell while keeping the other
quantities constant \citep[see also][]{Kay00}.

\subsection{High Mach-number problem}
\label{aHighMach}
In cosmological simulations, one often encounters flows of 
high velocity and low pressure. In these situations, the numerical 
computation of the difference $E - E_{\mathrm{kin}}$,
needed for the computation of the pressure, might might not yield 
reasonable results. This is known as the 
\emph{high Mach-number problem}. To overcome this problem we implement the 
algorithm suggested by \citet{Feng04}. In addition the conservative quantities we
also follow the evolution of a modified entropy density 
$S = p / \rho^{\gamma- 1}$. In high Mach flows, where $E \approx
E_{\mathrm{kin}}$, we use $S$ to compute the pressure, while elsewhere $E$ is
used. After the pressure is computed the quantity \emph{not} used for the
computation is recomputed using $p$, thus keeping both quantities synchronized.

\subsection{Hydrodynamic solver}

If we set the right-hand side of Eq. (\ref{eRho} - \ref{eS})
to zero we obtain the homogeneous Euler-equations. These equations are used to 
compute the pure hydrodynamic evolution of the fluid. This problem is solved
using the MUSCL-Hancock scheme as presented in \citet{Toro99}.
To ensure the monotonicity of the solution the MINMOD slope limiter is applied. 
The inter-cell fluxes are computed using a HLLC Riemann-solver. 
This solver approximates the analytic solution of the Riemann problem by three
waves separating (with velocities $S_L$, $S_\star$, $S_R$) four different
states: the left ($L$) and the right ($R$) initial state, and regions left
($\star L$) and right ($\star R$) of the contact discontinuity. We use the
algorithm given in \citet[chap. 10.4 and 10.5]{Toro99} and add a prescription
for the computation of the modified entropy density and, for non-IE
simulations, the number densities in the central regions $\star L$ and $\star R$:
\begin{eqnarray}
S_{\star K} &=& \rho_K \left(\frac{S_K - u_K}{S_K - S_\star}\right)
\frac{p_K}{\rho^{\gamma}} \\
n_{i , \star K} &=& 
\rho_K \left(\frac{S_K - u_K}{S_K -S_\star}\right)   
\frac{n_{i , \star K}}{\rho_K} \;,
\end{eqnarray}
where $K = L$ or $K = R$ for the left or right states. The wave
speeds are computed by
\begin{eqnarray}
S_L &=& \min\left[u_L - a_L,u_R - a_R\right]\\
S_R &=& \max\left[u_L + a_L,u_R + a_R\right]\\
S_\star &=& \frac{p_R - p_L + \rho_L u_L \left(S_L - u_L\right)
- \rho_R u_R \left(S_R - u_R\right)}
{\rho_L \left(S_L - u_L\right) - \rho_R \left(S_R - u_R\right)} \;.
\end{eqnarray}
Once the state on the interface is determined, the fluxes can be computed and
the conservative quantities can be updated.

\subsection{Gravitation}

The gravitational potential is obtained though the classical algorithm used 
by Particle-Mesh codes. We solve Poisson's equation with
Fourier-transformations and Green's function \citep{Hockney88}.
Then, the conservative quantities are updated
with the gravitational source terms. To perform the needed
Fourier-transformations we use the public available FFTW library \citep{FFTW}.

\subsection{Chemical evolution}

The evolution of the number densities and the heating and
cooling originate in the same physical processes and have similar timescales.
It is therefore necessary to compute their evolution in a similar way. In the
IE case a solution to $\Xi_i = 0$ can be found by iteration. The situation is
more difficult in the non-IE case. Here, the integration of the system of 
ordinary differential equations $\dot{n} = \Xi_i$ is performed.
These are stiff ordinary differential equations, and therefore most codes use 
implicit methods for their solution. In our code, we use a 
different approach and adopt the \emph{modified Patankar scheme} developed in 
biochemical oceanography \citep{Burchard03}. Although it is explicit it ensures
the positivity of temperature and number densities and conserves total amount
of hydrogen and helium. The modified Euler-Patankar scheme reads:
\begin{equation}\label{eModpatankar}
n^\prime_i = n_i + \Delta t \left( \sum_k p_{ik} \frac{n^\prime_k}{n_k} - 
\sum_k d_{ik} \frac{n^\prime_i}{n_i}\right) \;.
\end{equation}
where $p_{ik}$ is the production matrix containing the rates 
producing species $i$ from species $k$, while $d_{ik}$ is the destruction
matrix containing the rates that transform species $i$ into species $k$.
That implies $p_{ik} = d_{ki}$ and all diagonal coefficients are zero. We
obtain:
\begin{eqnarray}
p_{12} &=& \alpha_{\ion{H}{ii}} \, n_{\ion{H}{ii}} \, n_e  \;=\; d_{21} \\
p_{21} &=& \left(\beta_{\ion{H}{i}} \, n_e 
+ \gamma_{\ion{H}{i}} \right) n_{\ion{H}{i}} \;=\; d_{12} \\
p_{34} &=& \left( \alpha_{\ion{He}{ii}}  + \xi_{\ion{He}{ii}} \right)
n_{\ion{He}{ii}} \, n_e \;=\; d_{43} \\
p_{43} &=& \left(\beta_{\ion{He}{i}} \, n_e 
+ \gamma_{\ion{He}{i}} \right) n_{\ion{He}{i}} \;=\; d_{34} \\
p_{45} &=& \alpha_{\ion{He}{iii}} \, n_{\ion{He}{iii}} \, n_e \;=\; d_{54}\\
p_{54} &=& \left(\beta_{\ion{He}{ii}} \, n_e 
+ \gamma_{\ion{He}{ii}} \right) n_{\ion{He}{ii}}  \;=\; d_{45} \;.
\end{eqnarray}
The other components vanish. If we define
\begin{eqnarray}
P_{ik} = \frac{\Delta t \,p_{ik}}{n_k}
\quad\textrm{ and }\quad
D_i = \sum_k \frac{\Delta t \, d_{ik}}{n_i} \;.
\end{eqnarray}
Eq. (\ref{eModpatankar}) can be further performed to
\begin{equation}
n^\prime_i = n_i + \sum_k P_{ik} x_k - D_{i} n^\prime_i = \frac{n_i}{1 + D_{i}}
+ \sum_k \frac{P_{ik}}{1 + D_{i}} \, x_k \;.
\end{equation}
This is an easy solvable system of linear equations. Since the $2 \times 2$ block
matrix for hydrogen and $3 \times 3$ matrix for helium are not coupled, this
system can be solved for them independently.

In the IE as well as in the non-IE case the
update of the pressure is performed using original Patankar-Trick
\citep{Patankar80}:
\begin{equation}
	p^\prime = p + \Delta t \left( \Gamma - \Lambda \frac{p^\prime}{p}
	\right) = \frac{p + \Delta t \, \Gamma}{1 - \Lambda / p} \;.
\end{equation}

\subsection{Thermal conduction}

The algorithm for the thermal conduction is carried out similar to the
hydrodynamic scheme. After the thermal fluxes between the cells are computed,
those fluxes are used to update the energy density and modified entropy density.
Like the hydrodynamic scheme this is done in an unsplit fashion. In order to
compute the thermal flux, first the temperature gradient is computed:
\begin{equation}
  \left( \nabla_x T \right)_{i_x+1/2} = \frac{T_{i_x+1} - T_{i_x}}{\Delta x}
  \,,
\end{equation}
where $i_x+1/2$ denotes the position of the interface and $i_x$ and $i_x+1$ the
cell left and right of the interface. Then, the conduction coefficient
$\kappa$ and the fraction of mean free path and temperature $\lambda_e / T$ are
computed in cell $i_x$ and $i_x+1$ and then extrapolated to the interface
$i_x+1/2$ by simple averaging. The heat flux is then
\begin{equation}
  j_{i_x+1/2} = \kappa \; \nabla_x T \,
  \left( 1 + \frac{\lambda_e}{T} \, \nabla_x T\right)^{-1} \;,
\end{equation}
where all quantities are located at $i_x+1/2$. In a final step, we update energy 
density and modified entropy density according to
\begin{eqnarray}
E_{i_x}^\prime &=& E_{i_x} + \frac{\Delta x}{\Delta t} 
\left( j_{i_x+1/2} - j_{i_x-1/2}\right)\\
S_{i_x}^\prime &=& S_{i_x} + \frac{\Delta x}{\Delta t} 
\frac{\gamma - 1}{\rho^{\gamma-1}}
\left( j_{i_x+1/2} - j_{i_x-1/2}\right) \;.
\end{eqnarray}

\end{document}